\documentclass [article]{nature}

\usepackage{float}
\usepackage{units}
\usepackage{graphicx}
\usepackage{color}
\usepackage{epstopdf}

\usepackage{amsmath}
\usepackage{amsfonts}
\usepackage{amssymb}
\usepackage{dcolumn}
\usepackage{bm}
\def\onlinecite{\citeonline}

\title{Extremely large magnetoresistance and ultrahigh mobility in the topological Weyl semimetal NbP} 

\author{Chandra Shekhar$^1$, Ajaya K. Nayak$^1$, Yan Sun$^1$, Marcus Schmidt$^1$, Michael Nicklas$^1$, Inge Leermakers$^2$, Uli Zeitler$^2$, Yurii Skourski$^3$, Jochen Wosnitza$^3$,  Zhongkai Liu$^4$, Yulin Chen$^5$, Walter Schnelle$^1$, Horst Borrmann$^1$, Yuri Grin$^1$, Claudia Felser$^1$, \& Binghai Yan$^{1,6~\ast}$}

\date{\today}
\begin{document}
\maketitle

\begin{affiliations}
\item Max Planck Institute for Chemical Physics of Solids, 01187 Dresden, Germany
\item High Field Magnet Laboratory (HFML-EMFL), Radboud University, 
Toernooiveld 7, 6525 ED Nijmegen, The Netherlands
\item Dresden High Magnetic Field Laboratory (HLD-EMFL), Helmholtz-Zentrum Dresden-Rossendorf,
01328 Dresden, Germany
\item Diamond Light Source, Harwell Science and Innovation Campus, Fermi Ave, Didcot, Oxfordshire, OX11 0QX, UK
\item Physics Department, Oxford University, Oxford, OX1 3PU, UK
\item Max Planck Institute for the Physics of Complex Systems, 01187 Dresden, Germany
\end{affiliations}

\begin{abstract}

Recent experiments have revealed spectacular transport properties of conceptually simple semimetals. 
For example, normal semimetals (e.g., WTe$_2$)~\cite{Ali:2014bx} have started  a new trend to realize a large magnetoresistance, which is the change of electrical resistance by an external magnetic field.
Weyl semimetal (WSM)~\cite{Wan2011} is a topological semimetal with massless relativistic electrons as 
the three-dimensional analogue of graphene~\cite{novoselov2005two} and promises exotic transport properties and surface states~\cite{Turner:2013tf,Hosur:2013eb,Vafek:2014hl}, which are different from those of the famous topological insulators (TIs)~\cite{qi2011RMP, Hasan:2010ku}. 
In this letter, we choose to utilize NbP in magneto-transport experiments 
because its band structure is on assembly of a WSM~\cite{Weng:2014ue,Huang:2015uu} and a normal semimetal. 
Such a combination in NbP indeed leads to the observation of remarkable transport properties,
 an extremely large magnetoresistance of 850,000\% at 1.85~K  (250\% at room temperature) in a magnetic field of 9~T without any signs of saturation, and ultrahigh carrier mobility of 5$\times$10$^6$ cm$^2$ V$^{-1}$ s$^{-1}$ 
 accompanied by strong Shubnikov--de Hass (SdH) oscillations. 
NbP presents a unique example to consequent design the functionality of materials by combining the topological and conventional phases. 

\end{abstract}


A Weyl semimetal (WSM) is a three-dimensional (3D) analog of graphene, in which the conduction and valence bands cross near the Fermi energy. The band crossing point, the so-called Weyl point,  acts as a magnetic monopole (a singular point of Berry curvature) in momentum space and always comes in a pairs. Unusual transport properties and surface states such as Fermi arcs are predicted,  stimulating strong interest in realizing the WSM state in real materials\cite{Wan2011,Burkov:2011de,Xu:2011dy}. If the time-reversal and inversion symmetries are respected, a pair of Weyl points is degenerate in energy, forming another topological phase called Dirac semimetal~\cite{Wang:2012ds,Wang:2013is}. WSMs and Dirac semimetals usually exhibit very high mobilities, as observed in transport experiments (such as Cd$_3$As$_2$)~\cite{Liang:2014ev,Narayanan2014}. Generally, semimetals are new platforms to realise a huge magnetoresistance (MR)~\cite{Baibich:1988he,Binasch:1989hi}, an effect that has been pursued intensively in emerging materials in recent years, because of its significant application in state-of-the-art information technologies~\cite{Parkin:2003id}.
Electrical transport in a semimetal usually consists of two types of carriers (electrons and holes), leading to large MR when a magnetic field is applied with an electron--hole resonance~\cite{Singleton2001band,Ali:2014bx}. In a simple Hall effect setup, the transverse current carried by a particular type of carrier may be nonzero, although no net transverse current flows when the currents carried by the electrons and holes compensate for each other. These nonzero transverse currents will experience a Lorentz force caused by the magnetic field in the inverse-longitudinal direction.  Such a back flow of carriers eventually increases the apparent longitudinal resistance, resulting in dramatic MR that is much stronger than that in normal metals and semiconductors. Thus, it is crucial to obtain high purity samples to realize a balance between electrons and holes and a high carrier mobility ($\mu$) as well, both of which will enhance the MR effect.


Elemental Bi~\cite{Yang:1999ey} and WTe$_2$ (Ref.~\onlinecite{Ali:2014bx}) exhibit large MR as typical examples of semimetals, in which  electron and hole pockets coexist on the Fermi surface (Fig. 1a). 
There is a special type of semimetal whose conduction-band bottom and valence-band top touch the Fermi surface at the same point in momentum ($k$) space (Figs. 1b and 1c). Many such semimetals exhibit a high carrier mobility and relatively large MR with a linear dependence on the magnetic field, such as zero-gap topological-insulator silver chalcogenides ~\cite{Xu:1997df,wzhang2011} and Heusler compounds~\cite{chadov2010,Shekhar:2012ge,Yan:2014hv}, the Dirac semimetal Cd$_3$As$_2$ ($\mu= 9 \times 10^6$ cm$^2$ V$^{-1}$ s$^{-1}$ at 5 K, MR$=1500\%$ at 1.5 K and 14.5 T) ~\cite{Liang:2014ev,He:2014cf,Feng:2014tm,He:2014wn}, and the WSM TaAs ($\mu=5\times10^5$ cm$^2$ V$^{-1}$ s$^{-1}$ at 2 K, MR$= 5.4\times10^5~\%$ at 10 K and 9 T)~\cite{Zhang:2015wu}. The high mobility may originate from the linear or nearly linear energy dispersion. The unsaturated linear MR is interpreted as a classical effect due to the strong inhomogeneity in the carrier density ~\cite{Parish:2003bw} or as a quantum effect by the linear energy dispersion at the band touching point ~\cite{Abrikosov:1998be}. The semimetal NbP combines the main features of the WTe$_2$-type (showing extremely large MR) and Cd$_3$As$_2$-type (showing ultrahigh mobility) semimetals in the band structure (Fig. 1d), exhibiting hole pockets from normal quadratic bands and electron pockets from linear Weyl bands. As we will see, NbP exhibits an ultrahigh carrier mobility comparable to that of Cd$_3$As$_2$ and an extremely large MR surpassing that of WTe$_2$.


The single crystal of NbP used for the present study and the respective X-ray diffraction patterns are shown in Fig. 2a.
The crystal structure of NbP is non-centrosymmetric space group $I4_1 md$ (Fig. 2b). No indication of twinning was found in the diffraction experiments. Both atom types have the same coordination number of six, and the same coordination environment in form of a trigonal prism. A detailed overview of the structural characterization is presented in the supplementary information.   A measurement of the temperature dependence of the resistivity, $\rho_{xx} (T)$, is a simple way to identify the electronic states of a material.  On the basis of our high-quality single crystals of NbP grown via chemical vapor transport reactions, $\rho_{xx} (T)$ is measured under various transverse  magnetic fields ranging from 0 to 9~T, as shown in Fig. 2c. At zero field, we observe  metallic behavior with $\rho_{xx}$ (300~K) $=73~\mu\Omega$~cm and a residual resistivity $\rho_{xx} (2~$K$)=0.63~\mu\Omega$~cm.  This results in a residual resistivity ratio [$\rho_{xx} (300~$K$)/\rho_{xx} (2~$K$)]=115$, which is directly related to the metallicity and quality of the crystal. 
Compared to other similar materials at low temperature (2~K), NbP exhibits a resistivity that is about 30 times lower than that of WTe$_2$ (Ref. \onlinecite {Ali:2014bx}) but 30 times higher than that of Cd$_3$As$_2$ (Ref.~\onlinecite{Liang:2014ev}).
NbP does not become superconducting for temperatures above 0.10~K. After applying magnetic fields, we observe a remarkable change in the resistivity. $\rho_{xx}(T)$ changes from positive slope (metallic) to negative slope (semiconducting) from a very small field of 0.1~T and becomes completely semiconducting at a field of 2~T. This may be due to the opening of a gap at the Weyl point. In general, a conventional semimetal does not display such behavior, whereas some small-gap or gapless semimetals (e.g., WTe$_2$\cite{Ali:2014bx} and Cd$_3$As$_2$\cite{Liang:2014ev}) exhibit a similar trend, usually at very high fields and low temperature.
Another important fact observed in the present material is that $\rho_{xx}(T)$ also increases dramatically due to the application of magnetic fields at room temperature (300 K).

We now focus on the MR measurement in NbP.  The MR is commonly calculated as the ratio of the change in the resistivity due to the applied  magnetic field ($H$), $[(\rho(H)-\rho(0))/\rho(0)]\times 100 \%$. Figure 2d displays the MR measured in transverse magnetic fields up to 9~T at different temperatures. At low temperatures, we find that NbP exhibits an extremely large MR $=8.5\times 10^5$~\% at 1.85~K in a field of 9~T. This MR is five times larger than that measured for the same field in WTe$_2$\cite{Ali:2014bx} and nearly two times as large as than that of TaAs~\cite{Zhang:2015wu}, another WSM predicted in the same family as NbP~\cite{Weng:2014ue,Huang:2015uu}. Upon increasing temperature, the MR of NbP remains almost unchanged up to 20~K and then starts to decrease at higher temperatures. We note that the MR is still as high as 250 \% in 9~T at room temperature (inset of Fig. 2d) . We have also measured the angular dependence of the MR on the direction of magnetic field.
Figure 2f shows the MR observed at different angles ($\theta$) between the field and current directions. 
The MR decreases slightly from $8.5\times 10^5 ~\%$ at $\theta=90^\circ$ (transverse) to $2.5\times 10^5$\% at $\theta=0^\circ$ (longitudinal). Thus, the MR varies only by a factor of 3.4, implying the relatively isotropic nature of the material, i.e., compared to the layered semimetal WTe$_2$.

A large MR is usually associated with a high mobility. The carrier mobility and concentration are two important parameters of a material, which can be derived from the Hall coefficient. We have performed Hall effect measurements in both temperature-sweep and field-sweep modes to improve the accuracy of our data. The field dependence of the Hall resistivity $\rho_{xy}(H)$ exhibits a linear characteristic at high fields (see Supplementary Information). However,  the nonlinear behavior in low fields indicates the involvement of more than one type of charge carriers in the transport properties. As seen from the inset of Fig. 2e, NbP exhibits a negative Hall coefficient, $R_H(T)$, up to 125~K which changes sign for temperatures above 125~K. For the sake of simplicity, we use the single-carrier Drude band model, $n_{e,h}(T)=1/[e R_H(T)]$, to calculate the carrier density and $\mu_{e,h}(T)= R_H(T)/\rho_{xx}(T)$ in order to estimate the mobility, where $n_e (n_h)$ and $\mu_e (\mu_h)$ are charge density and mobility of the electron (hole), respectively. We use the slope of $\rho_{xy}(H)$ at high fields to calculate the Hall coefficient (Fig. 2e). The electron carrier concentration, $n_e$, is found to be $1.5\times10^{18}$ cm$^{-3}$ at 2~K and increases  slowly with temperature, exhibiting a semimetal-like or very small gap-like behavior. The mobility plays a major role in the charge transport in a material and consequently determines the efficiency of various devices. Here, NbP exhibits an ultrahigh mobility of $5\times 10^6$cm$^2$V$^{-1}$s$^{-1}$ at 2~K.  
This value is close to that of Cd$_3$As$_2$~\cite{Liang:2014ev}, which is reported to exhibit a mobility of $9\times 10^6$ cm$^2$ V$^{-1}$s$^{-1}$, one order magnitude higher than that of TaAs at 2~K. Furthermore, it  has been shown that the mobility in Cd$_3$As$_2$ scales with the residual resistivity. Hence, it can be proposed that materials with a low residual resistivity exhibit high mobilities, with the present case being a good example.


The low field measurements (Fig. 2d and 2f) do not show any saturation of the MR up to 9~T in the entire temperature range. In order to pursue this finding to even higher fields, we have performed transverse MR measurements up to 30~T in dc magnetic fields as shown in Fig. 3a. The MR increases up to $3.6\times 10^6$\% for a field of 30~T at 1.3~K and still shows no tendency of saturation. We have further corroborated this trend of a non-saturating magneto-resistance by performing experiments in pulsed magnetic fields up to 62~T at 1.5~K (inset of Fig 3a). The MR reaches a value of $8.1\times 10^6$ without any indication for saturation. The MR continues increasing up to 62T the maximum field reached in the pulsed magnetic field experiments. At 62T we find an MR of $8.1\times 10^6$. As already indicated in the 9~T MR data depicted in Fig. 2d and 2f, SdH quantum oscillations appear for $T\leq30$~K. At the lowest temperature (1.85~K) the oscillations start for fields as low as 1~T. Since SdH oscillations only appear when the energy spacing between two Landau levels is larger than their broadening due to disorder, we can estimate a lower limit to the quantum mobility of the carriers involved $\mu_q > 10^5$ cm$^2$V$^{-1}$s$^{-1}$, in agreement with the large electron (transport) mobility extracted from the Hall effect.

The physical parameters of the charge carriers are derived from the Fermi surface that is measured by the oscillations observed in the transport properties. Both $\rho_{xx}$ and $\rho_{xy}$ measured up to 9~T exhibit very clear SdH oscillations starting from 1~T. This indicates a very low effective mass, resulting in a high mobility. To obtain the amplitude of the SdH oscillations, $\Delta \rho_{xx}$ from $\rho_{xx}$, we subtracted a smooth background. The results are plotted as a function of 1/($\mu_0H$) at various temperatures in Fig. 3b. As expected, the oscillations are periodic in 1/($\mu_0H$).  They occur due to the quantization of energy levels which directly gives the effective mass of the charge carriers. A single oscillation frequency $F=$ 32~T is identified from the data shown in Fig. 3b, which corresponds to $1/(\mu_0 \Delta H) = 0.03127$~T$^{-1}$.
We calculated a corresponding cross-sectional area of the Fermi surface $A_F=0.003~{\rm \AA}^{-2}$, from the Onsager relationship $F=(\Phi_0/2\pi^2)A_F$, where $\Phi_0$ is the single magnetic flux quantum. Supposing a circular cross section, a very small Fermi momentum $k_F=0.0312~{\rm \AA}^{-1}$ is obtained by assuming a circular cross section. The cyclotron effective mass of the carriers is determined by fitting the temperature dependence of amplitude of the oscillations to the Lifshitz-Kosevich formula,

\[\frac{\Delta\rho_{xx}(T,B)}{\rho_{xx}(0)} = e^{-2\pi^2k_BT_D/\beta}\frac{2\pi^2k_BT/\beta}{sinh(2\pi^2k_BT/\beta)}~~,\]

where $\beta=ehB/2\pi m^*$ and $T_D = h/4\pi^2\tau k_B$ are the fitting parameters, which directly result in the effective mass $m^*$ and quantum lifetime $\tau$ of the charge carriers. The value of $m^*$ and $\tau$ calculated from the temperature dependent SdH oscillations in 8.2~T are 0.076~$m_0$ and $1.5\times 10^{-12}$~s, respectively, where $m_0$ is the bare mass of the electron. These values are comparable to that reported for Cd$_3$As$_2$ ~\cite{Liang:2014ev}. We also find a very large Fermi velocity $v_F$ = $h k_F/(2\pi m^*)$ = 0.48$\times 10^6$~m~s$^{-1}$. The large Fermi velocity and low effective mass are responsible for the observed ultra-high mobility in NbP.

The SdH oscillations at different temperatures obtained from the 30~T dc magnetic field measurements are plotted in Fig. 3c.
At the lowest temperature, marked by arrows, the SdH maxima at 8~T (0.125 T-1) and 11~T(0.091 T-1) start to split into two distinct peaks, and the maximum at 16~T (0.0625) develops into four peaks. This can be assigned to the lifting of the spin-degeneracy and the degeneracy of the dual Weyl point.


To further understand the transport properties of NbP, we have performed \textit{ab-initio} band-structure calculations.
NbP crystallizes in a body-centered-tetragonal lattice with the nonsymmorphic space group $I4_1 md$. The lack of inversion symmetry of the lattice leads to the lifting of spin degeneracy in the band structure (Fig. 4a). Near the Fermi energy, twelve pairs of Weyl points lie aside the central planes in the Brillouin zone (Fig. 4b),  consistent with recent calculations\cite{Weng:2014ue,Huang:2015uu}. An important feature in the band structure is that the Fermi energy crosses the quadratic-type valence bands and also the linear Weyl-type conduction bands, leading to hole and electron pockets on the Fermi surface (Fig. 4c), respectively. Four equivalent hole pockets appear around the $Z$ point. Four larger and eight smaller electron pockets exist near the $\Sigma$ and $N$ points, respectively, because the four pairs of Weyl points near $\Sigma$ are lower in energy than the other eight pairs near $N$. 
The distribution of the carrier pockets is relatively isotropic compared to the layered material WTe$_2$, for example, explaining the weak anisotropy of the observed MR. 

As we have demonstrated here, NbP is an exotic semimetal with interesting transport properties. 
As seen in the band structure, it combines the electronic structures of a normal semimetal and a Weyl semimetal together.
Similar to normal semimetals such as WTe$_2$ and Bi, NbP exhibits both electron and hole pockets at different positions in the Brillouin zone; however unlike WTe$_2$ and Bi, its electron pockets are relevant to the linear Weyl bands, from which the high mobility of the high-quality samples may originate. In contrast to  Weyl semimetals (e.g., TaAs) and Dirac semimetals (e.g., Cd$_3$As$_2$), in which the Fermi energy may cross only one type of bands (electron or hole),
NbP naturally hosts both types of carriers and consequently exhibits a huge MR in an electron-hole resonance situation.



\paragraph{Methods}
High-quality single crystals of NbP were grown via a chemical vapor transport reaction using iodine as a transport agent. 
Initially, polycrystalline powder of NbP was synthesized by a direct reaction of niobium (Chempur 99.9\%) and red phosphorus (Heraeus 99.999\%) kept in an evacuated fused silica tube for 48 hours at 800 $^{\circ}\mathrm{C}$. Starting from this microcrystalline powder, the single-crystals of  NbP were synthesized by chemical vapor transport in a temperature gradient starting from 850 $^{\circ}\mathrm{C}$ (source) to 950 $^{\circ}\mathrm{C}$ (sink) and a transport agent  with a concentration of 13,5 mg/cm$^3$ iodine (Alfa Aesar 99,998\%)~\cite{martin1988chemischen}. The orientation and crystal structure of the present single crystal were investigated using the diffraction data sets collected on a Rigaku AFC7 diffractometer equipped with a Saturn 724+ CCD detector (monochromatic Mo$K_{\alpha}$ radiation, $\lambda =$ 0.71073~\AA). Structure refinement was performed by full-matrix least-squares on $F$ within the program package WinCSD.

The transport measurements were performed in different physical property measurement systems (PPMS, Quantum Design, ACT option, home build adiabatic demagnetization stage). The 30~T static magnetic-field measurements were performed at the High Field Magnet Laboratory HFML-RU/FOM in Nijmegen, and the pulsed magnetic field experiments were carried out at the Dresden High Magnetic Field Laboratory HLD-HZDR; both laboratories are member of the European Magnetic Field Laboratory (EMFL).
The \textit{ab-initio} calculations were performed within the framework of density-functional theory (DFT) with the generalized gradient approximation~\cite{perdew1996}.  We employed the Vienna $ab-initio$ simulation package with a plane-wave basis~\cite{kresse1996}. The core electrons were represented by the projector-augmented-wave potential. For calculation of the  Fermi surface, we have used the $ab-initio$ tight-binding method based on maximally localized Wannier functions (MLWFs)~\cite{Mostofi2008}. \\

{\bf Acknowledgements}  This work  was financially supported  by the  Deutsche Forschungsgemeinschaft
DFG  (Project No.EB 518/1-1 of DFG-SPP 1666 "Topological Insulators") and by the ERC Advanced Grant No. (291472) "Idea Heusler". We acknowledge the support of the High Magnetic Field Laboratory Dresden (HLD) at HZDR and High Field Magnet Laboratory Nijmegen (HFML-RU/FOM),  members of the European Magnetic Field Laboratory (EMFL). 

\begin{addendum}
 \item[Competing Interests] The authors declare that they have no
competing financial interests.

{\bf Author Contributions} All authors contributed substantially to this work.

 \item[Correspondence] Correspondence and requests for materials
should be addressed to B.Yan~(email: yan@cpfs.mpg.de).
\end{addendum}



\begin{figure}
\includegraphics[angle=0,width=14cm,clip]{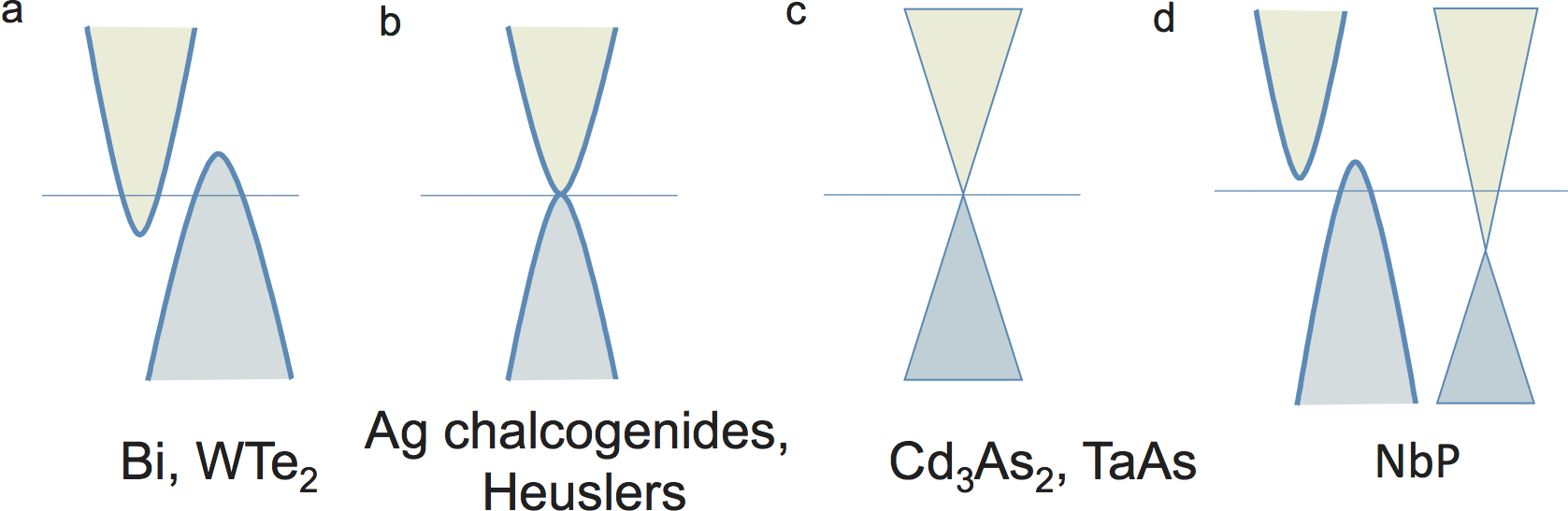}
\caption{{\bf\boldmath Band structure for different semimetals.} Schematic illustration of different types of semimetals and representative materials.  {\bf a}, Normal semimetal with the coexistence of electron and hole pockets. 
{\bf b}, Semimetal with quadratic conduction and valence bands touching at the same momentum point. {\bf c}, Weyl semimetal and Dirac semimetal. {\bf d}, Semimetal with one hole pocket from the normal quadratic band and one electron pocket from the linear Weyl semimetal band for  NbP. The valence and conduction bands are indicated by bluish and greenish shading, respectively. The Fermi energy is marked by horizontal lines.}
\end{figure}


\begin{figure}
\includegraphics[angle=0,width=14cm,clip]{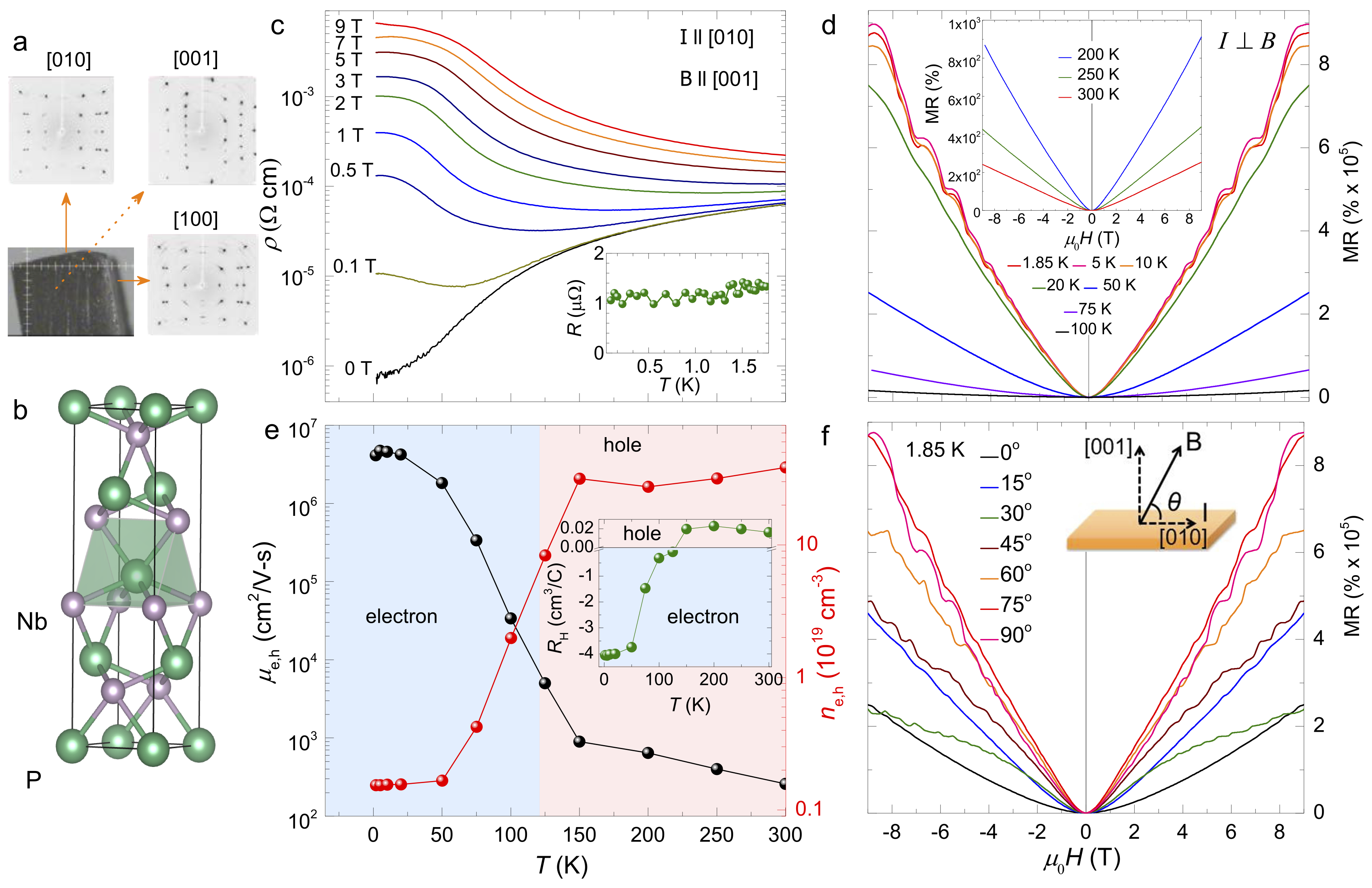}
\caption{{\bf\boldmath Crystal structure, magnetoresistance and mobility.} {\bf a}, Orientation of the measured single crystal NbP with the respective X-ray diffraction patterns. {\bf b}, Crystal structure of NbP in a body-centered-tetragonal lattice. {\bf c}, Temperature dependence of the resistivity measured at different transverse magnetic fields displayed next to the corresponding curve. The inset of {\bf c} shows the temperature dependence of resistance measured up to 0.1~K in zero field.  {\bf d}, Transverse magnetoresistance measured at different temperatures with field up to 9~T. The inset shows the magnetoresistance at higher temperatures. {\bf e}, Temperature dependencies of the mobility (left ordinate) and carrier density (right ordinate). The inset shows the evolution of the Hall coefficient with temperature. The temperature regimes where electrons and holes  act as main charge carriers are marked with bluish and reddish colors, respectively. {\bf f}, Magnetoresistance measured at different angles, $\theta$, between the current and the magnetic field as shown schematically in the inset.}
\end{figure}


\begin{figure}
\includegraphics[angle=0,width=14cm,clip]{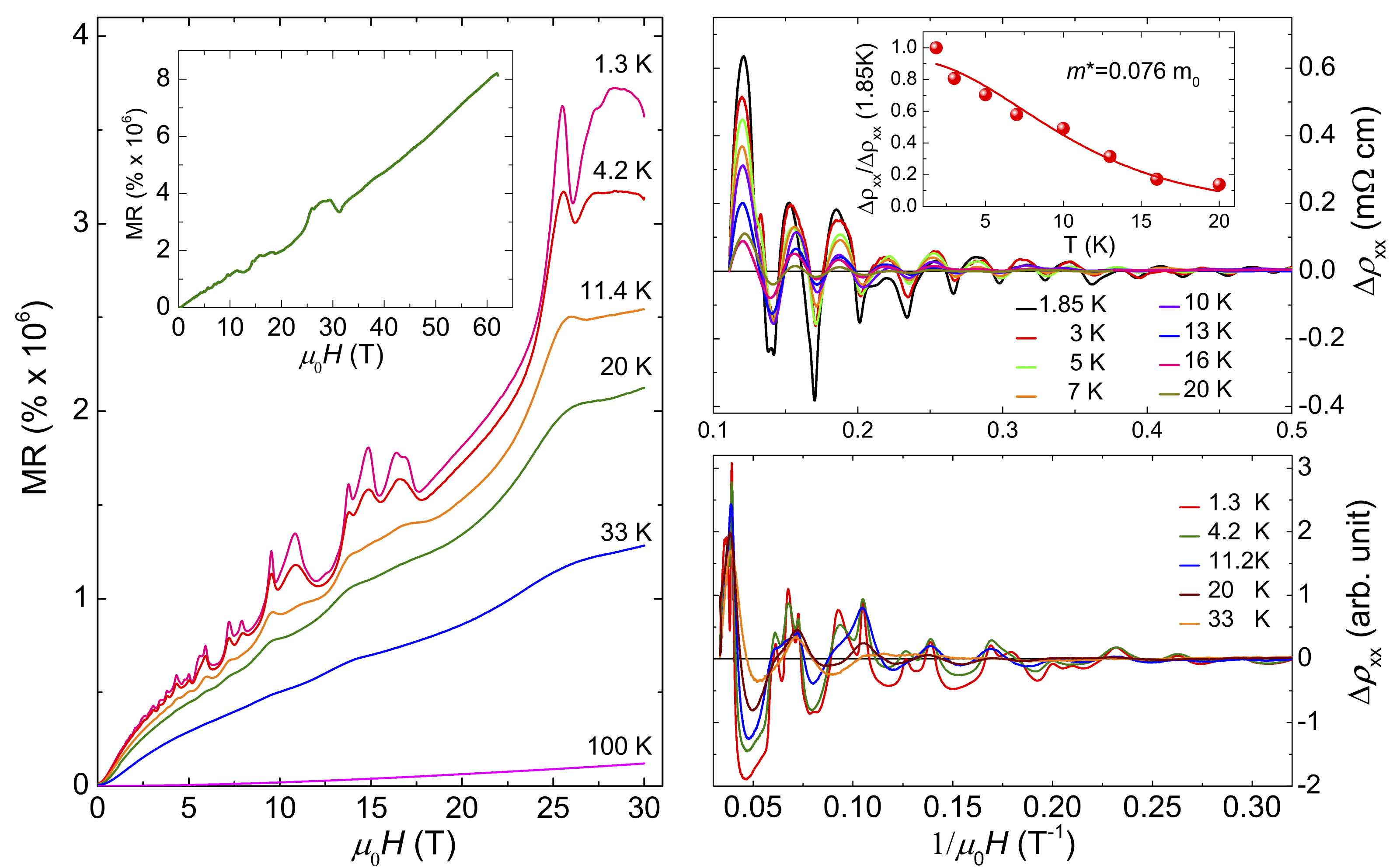}
\caption{{\bf\boldmath High-field magnetoresistance and SdH oscillation.} {\bf a}, Transverse MR measured up to 30~T static magnetic field at different temperatures. The inset of {\bf a} shows MR measured up to 62~T in a pulsed magnetic field.  {\bf b},  SdH oscillations after subtracting the background from the 9~T $\rho_{xx}$ measurements. The inset of {\bf b} shows the temperature dependence of the relative amplitude of $\Delta\rho_{xx}$ for the SdH oscillation at 8.2~T. The solid line is a
fit to the Lifshitz-Kosevich formula. {\bf c}, SdH oscillations after subtracting the background from the 30~T $\rho_{xx}$ measurements. }
\end{figure}


\begin{figure}
\includegraphics[angle=0,width=14cm,clip]{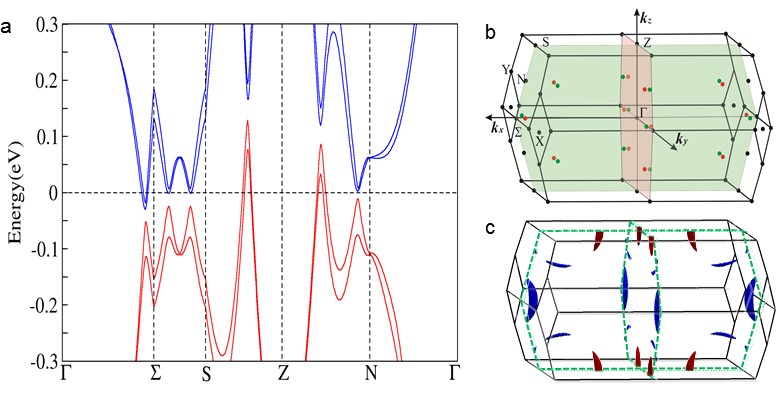}
\caption{{\bf\boldmath Structure and electronic properties of NbP.} {\bf a}, $Ab-initio$ band structure. The Fermi energy is shifted to zero. The valence and conduction bands are indicated by the red and blue lines, respectively. {\bf b}, The first Brillouin zone in momentum space for the primitive unit cell. Green and red dots represent the Weyl points with opposite chirality. {\bf c}, Fermi surfaces. The blue and red surfaces represent electron and hole pockets, respectively. }
\end{figure}


\end{document}